\documentclass[prd,superscriptaddress,floatfix,nofootinbib]{revtex4-2}
\usepackage[colorlinks,linkcolor=blue,anchorcolor=violet,citecolor=red]{hyperref}
\usepackage{amsmath,amssymb}
\usepackage{graphicx}
\usepackage{float}
\usepackage{comment}
\usepackage{xcolor}
\usepackage{bm,latexsym,amsmath,amssymb,amsfonts,mathrsfs}
\usepackage{bbold}
\usepackage{ascmac}
\usepackage{physics}
\usepackage{color,float}
\usepackage{enumerate}
\usepackage{bmpsize}

\usepackage{bbm}
\usepackage{CJKutf8}
\usepackage{ulem}

\newcommand{\apjl}{Astrophys. J. Lett. }

\newcommand{\araa}{Ann. Rev. Astron. Astrophy. }
\newcommand{\mnras}{Mon. Not. Roy. Astron. Soc. }

\newcommand{\nphysa}{Nucl. Phys. A}

\begin{document}
\title{
\large 
On the Nucleon Effective Mass in Neutron Stars Cooling 
}

\author{Yuuki Sugiyama}
 \email{sugiyama.yuki@issp.u-tokyo.ac.jp}
\affiliation{Institute for Solid State Physics, the University of Tokyo, Kashiwa, Chiba 277-8581, Japan}
 
\author{Akira Dohi}
 \email{akira.dohi@riken.jp}
\affiliation{Interdisciplinary Theoretical and Mathematical Sciences (iTHEMS) Center, RIKEN, Wako, Saitama 351-0198, Japan}

\begin{abstract}

The cooling of neutrons stars, which are composed of ultra-high-density nuclear matter, strongly depends on the equation of state because the dominant cooling sources are neutrinos produced in their interiors. 
Here, we investigate the cooling properties of neutron stars, focusing on effective nucleon masses, whose impacts have been recently examined in supernova explosion and proto-neutron star cooling.
The uncertainties of effective masses originate from the difference in the attractive interaction of scalar meson and the definition of the Dirac and Landau masses.
At first, we employ so-called Walecka models that allow these two masses to be introduced within a unified framework and evaluate the resulting cooling curves.
Based on the Walecka model, we compute the emissivity of the neutrons and the direct Urca process and their cooling curves.
We demonstrate that the Dirac and Landau masses affect cooling of massive neutron stars, although most previous cooling simulations do not distinguish them. 
We also perform cooling simulations with two realistic equation of states based on relativistic mean-field (RMF) theory, whose differences arise from properties of only effective masses among characteristic parameters.
These results are broadly consistent with those obtained from the Walecka models, indicating that the treatment of effective nucleon masses may be an important source of uncertainty in neutron-star cooling.
\end{abstract}

\maketitle

\section{Introduction}
The behavior of matter under extreme conditions remains one of the central questions in modern physics. 
Neutron stars (NSs)  offer a particularly important realization of such conditions, where matter is compressed to high densities~\cite{Sinha2026}.
After Type-II supernova explosion of massive stars with the mass 
$\sim (8-20)~M_\odot$, 
newborn(proto) NSs cool down through neutrino emission from their cores and eventually evolve into cold NSs.
This neutrino cooling era lasts for 
$t\lesssim10^5~{\rm yr}$, 
after that the photon cooling becomes dominant. 
From the observed age and surface temperature/photon luminosity of isolated NSs, one can extract information on the interior of NSs, in particular the equation of state (EOS) models.
Unlike well-used mass-radius observations, such cooling observations enable us to constrain the compositions and quantum states of NS matter, not only its stiffness. 
Since the first work done by Tsuruta \& Chamelon \cite{1966CaJPh..44.1863T}, many people have shown the usefulness of cooling theory as a tool to probe high-density matter, in particular for the impacts of rapid cooling and nucleon superfluidity (for reviews, see \cite{2004ARA&A..42..169Y,2006ARNPS..56..327P,2021PrPNP.12003879B}). 

The neutrino emissivities of particle reactions inside the core generally depend on particle fraction, the strength of superfluidity, and nucleon effective masses of relevant nuclei, in addition to density and temperature themselves. 
For the former two aspects, there are numerous studies to investigate each alone in cooling curves of isolated NSs. 
On the other hand, there are few studies to investigate the effective mass in cooling curves, at least alone. 
Although the experimental studies to constrain nucleon effective masses have been done (for a review, see \cite{2018PrPNP..99...29L}), it is difficult for any available nuclear experiments to measure the effective masses in high-density regions. 
Put differently, uncertainties of effective masses in high-density matter might always drastically change the fate of NS evolution.

Recently, the impacts of effective masses have attracted considerable attention in high-energy astrophysical phenomena. 
For example, the fate of gravitationally collapsed massive stars, characterized by the strength of the supernova explosion, is shown to be highly changed by effective-mass uncertainties~\citep{2019PhRvC.100e5802S,2020PhRvL.124i2701Y}.
The subsequent cooling of proto NSs could also be affected, which may be imprinted in observed supernova neutrino luminosity~\citep{2019ApJ...878...25N}. 
According to these findings on the decisive effects, the unified EOS models with changing effective mass at saturation density, while keeping other characteristics on energy per baryon (such as parameters of symmetry energy), have been developed~\citep{2025ApJ...980...54L}, which has been recently applied for studying compositions inside supernovae~\citep{2026arXiv260411431M}. 
In this work, we focus on the subsequent long-term cooling after the proto-NS cooling phase as a site to see the uncertainties of effective masses.  

In cold NSs, it is well known that the effective mass could play a role in cooling curves via neutrino emissivity, thermal conductivity, and specific heat. 
In the Skyrme interaction model, the uncertainties of the effective mass play a minor role in cooling curves under the Skyrme EFD model \cite{2020IJMPE..2930007K}, which may not hold in other frameworks. 
In particular, the relativistic framework gives a different definition of the effective mass between the Dirac and Landau masses, the latter of which can also be defined in a non-relativistic manner as in the Skyrme nuclear model.

In the present paper, we consider a 
$(3+1)$-dimensional 
Walecka model to clarify the difference between the Dirac and Landau masses within a unified framework based on quantum field theory.
The Walecka model, or the 
$\sigma$-$\omega$ 
model, 
is a relativistic effective model of nuclear matter in which nucleons interact through scalar
($\sigma$)
and vector
($\omega$)
meson exchange~\cite{Walecka1974, Walecka2004, Glendenning1997}.
Because of its simplicity and its ability to describe dense baryonic matter at the mean-field level, it has been extensively used in studies of the EOS models relevant to neutron-star matter. 
In the Walecka model, the Dirac mass is defined as the effective mass of a baryon particle modified by the scalar mean-field.
In contrast, the Landau mass characterizes the quasiparticle dispersion near the Fermi surface and is determined by the Fermi velocity.

In this work, we derive the Dirac and Landau masses in the low-temperature approximation and obtain analytic expressions useful for evaluating neutron-star cooling curves. 
Within the Walecka model, the Dirac and Landau masses exhibit similar density dependence at low densities. 
At high densities, however, the Dirac effective mass decreases while the Landau mass increases, reflecting the different physical quantities encoded in their definitions.
We then compute cooling curves based on the field-theoretic approach for the neutron-neutron bremsstrahlung and direct Urca processes, using two choices of coupling constants. 
The resulting curves show different behavior in the high-density regime, reflecting the difference between the Dirac and Landau masses. 
We further compute cooling curves in a more realistic setup and compare them with those obtained from the field-theoretic approach.
Our analysis shows that uncertainties associated with the different definitions of effective mass and the attractive scalar interactions affect surface temperature at an order of a few $\mathcal{O}(10^{-2})$.
These results indicate that uncertainties in the effective masses provide a non-negligible contribution to the systematic uncertainty in neutron-star cooling calculations. 
They should therefore be included when cooling observations are used to constrain the properties of dense matter.
 
This paper is organized as follows.
In Sec. II, we introduce the Dirac and Landau masses based on a RMF theory in finite temperature.
In particular, we show the behaviors of the Dirac and Landau masses for zero and finite temperature cases. 
In Sec. III, we discuss the importance of the difference between the Dirac and Landau masses on cooling curves in our formalism.
To validate our discussion based on the Walecka model, we also compute cooling curves using two realistic EOS models that differ only in their properties of the effective masses.
Sec.~IV is devoted to the conclusion.
In Appendix A, we summarize the derivation of the grand potential 
$\Omega$ 
in finite temperature.
Throughout the present paper, we use the natural units 
$c=\hbar=k_{\text{B}}=1$
and adopt the metric convention 
$\eta_{\mu\nu}=\mathrm{diag}(+1,-1,-1,-1)$ 
in 
$(3+1)$-dimensional 
flat spacetime.

\section{Dirac mass and Landau mass}
In this section, we formulate the finite-temperature Walecka model within the relativistic mean-field approximation and introduce the thermodynamic and quasiparticle quantities relevant to the low-temperature regime.
After defining the Dirac and Landau masses, we derive the energy density, entropy density, and specific heat and evaluate their leading low-temperature behavior at fixed baryon density. 
Our main goal is to demonstrate that the Dirac and Landau masses are generally distinct and to examine their dependence on density.

\subsection{Walecka model in finite temperature}
Here, we outline the RMF treatment of the Walecka model and introduce the Dirac mass and the Landau mass.
The Lagrangian density of the Walecka model in 
$3+1$-dimensions 
is given by
\begin{align}
\mathscr{L}
=
\frac{1}{2}\partial_{\mu}\sigma \partial^{\mu}\sigma
-\frac{1}{2}m^2_{\sigma}\sigma^2
-\frac{1}{4}F_{\mu\nu}F^{\mu\nu}
+\frac{1}{2}m^2_{\omega}\omega_{\mu}\omega^{\mu}
+\bar{\psi}
\left(
i\gamma^{\mu}\partial_{\mu}-
m_{\psi}+g_{\sigma}\sigma-
g_{\omega}\gamma^{\mu}\omega_{\mu}
\right)
\psi,
\end{align}
where the fields
$\sigma$ 
and 
$\omega_{\mu}$
denote the scalar and the vector meson with masses
$m_{\sigma}$
and 
$m_{\omega}$, respectively.
The 
$\sigma$ ($\omega$ ) field interacts with the nucleon field
$\psi$
with a free nucleon mass
$m_{\psi}$
through the coupling constant
$g_{\sigma}$ ($g_{\omega}$).
The gamma matrices
$\gamma^{\mu}$
satisfy the anti-commutation relation
$\{\gamma^{\mu}, \gamma^{\nu}\}=2\eta^{\mu\nu}$.
The anti-symmetric tensor 
$F_{\mu\nu}$
is defined as
$F_{\mu\nu}
=\partial_{\mu}\omega_{\nu}-\partial_{\nu}\omega_{\mu}$.

To describe the nucleon dynamics, we adopt the mean-field approximation for the meson fields. 
The mean-field approximation neglects the fluctuation effect of meson fields and treats them as constant value.
Assuming homogeneous and isotropic nuclear matter, we replace the meson fields by their spacetime-independent expectation values,
\begin{align}
\sigma(t,\bm{x})
\to
\sigma,
\quad
\omega_{\mu}(t, \bm{x})
\to
\omega \delta_{\mu,0}.
\end{align}
The spatial components of 
$\omega_{\mu}$
vanish as a consequence of rotational symmetry~\cite{Glendenning1997, Papazoglou1997, Kapusta2006}.
Thanks to the mean-field approximation, the Lagrangian density reduces to the following form
\begin{align}
\mathscr{L}
=
-\frac{1}{2}m^2_{\sigma}\sigma^2
+\frac{1}{2}m^2_{\omega}\omega^2
+\bar{\psi}
\left(
i\gamma^{\mu}\partial_{\mu}-m_{\psi}
+g_{\sigma}\sigma-g_{\omega}\gamma^{0}\omega
\right)
\psi.
\end{align}

To describe the thermal evolution of nuclear matter in the cooling equation introduced in the next section, we need the specific heat (per unit volume) and neutrino emissivities. 
Their evaluation requires the grand potential and the associated thermodynamic quantities, including the entropy density.
We therefore formulate the finite-temperature thermodynamics of the Walecka model using the imaginary-time formalism 
$t=-i\tau$
to incorporate finite temperature and finite density~\cite{Kapusta2006, Cavagnoli2010}.
The imaginary time formalism leads to anti-periodicity for the fermion field
$\psi(\tau+\beta, \bm{x})
=-\psi(\tau, \bm{x})$
with the inverse temperature 
$\beta=1/T$.
The finite density of the nucleon
$N= \psi^{\dagger}\psi$
is introduced by considering the partition function 
$Z$ 
as
\begin{align}
Z=
\int \mathcal{D}\bar{\psi}
\mathcal{D}\psi
e^{-S[\bar{\psi}, \psi]},
\label{partition}
\end{align}
where the Euclidean action 
$S[\bar{\psi}, \psi]$
is
\begin{align}
S[\bar{\psi}, \psi]
&=
\int_{0}^{\beta} d\tau 
\int d^3x
\left[
\frac{1}{2}m^2_{\sigma}\sigma^2
-\frac{1}{2}m^2_{\omega}\omega^2
+\bar{\psi}
\left(
\gamma^{0}\partial_{\tau}
-i\gamma^{i}\partial_{i}+m_{\psi}
-g_{\sigma}\sigma+g_{\omega}\gamma^{0}\omega
\right)
\psi
-\mu N
\right]
\nonumber\\
\quad
&=
\int_{0}^{\beta} d\tau 
\int d^3x
\left[
\frac{1}{2}m^2_{\sigma}\sigma^2
-\frac{1}{2}m^2_{\omega}\omega^2
+\bar{\psi}
\left[
\gamma^{0}
\left(
\partial_{\tau}-\mu^{*}
\right)
-i\gamma^{i}\partial_{i}+m_{\text{D}}
\right]
\psi
\right].
\end{align}
In the second line, we introduced the Dirac mass (effective baryon mass) 
$m_{\text{D}}
=m_{\psi}-g_{\sigma}\sigma$
and the effective chemical potential
$\mu^{*}
=\mu-g_{\omega}\omega$.
The scalar and vector mean fields modify the single-particle Dirac equation by shifting the nucleon mass and chemical potential, respectively, thereby describing the nucleon as a quasiparticle in nuclear matter.
These shifts arise from the scalar and vector components of the nucleon self-energy, 
$g_{\sigma}\sigma$
and
$g_{\omega}\omega$,
respectively.
\footnote{
In the simplest mean-field treatment, the vector mean field shifts the single-particle energy by a constant and therefore does not directly contribute to either the Dirac mass or the Landau mass.
This is no longer necessarily true in extended RMF models with momentum-dependent self-energies. 
For instance, Typel introduced derivative meson-nucleon couplings that generate momentum-dependent scalar and vector self-energies, allowing the vector self-energy to influence the quasiparticle dispersion and thus the Landau mass~\cite{Typel2005,Typel2010}.
}

The grand potential 
$\Omega$ 
is defined by the partition function
$\Omega
=-\log{Z}/\beta V$ 
with the volume of the system
$V=\int d^3x$.
After integrating out the fermion fields 
$\psi$ and $\bar{\psi}$, 
we obtain (e.g., detailed derivation, see Refs.~\cite{Kapusta2006, Cavagnoli2010, Schmitt2010} or Appendix A)
\begin{align}
\Omega
&=
\frac{1}{2}m^2_{\sigma}\sigma^2
-\frac{1}{2}m^2_{\omega}\omega^2
-\frac{1}{\beta}
\sum_{n}
\int \frac{d^3k}{(2\pi)^3}
\text{tr}_{\text {D}}
\log
\left[
G^{-1}(i\omega_{n}, \bm{k})
\right]
\nonumber\\
\quad
&=
\frac{1}{2}m^2_{\sigma}\sigma^2
-\frac{1}{2}m^2_{\omega}\omega^2
+\frac{2}{\beta}
\int \frac{d^3k}{(2\pi)^3}
\left[
\log
\left[
1-f_{-}(\varepsilon_{k})
\right]
+
\log
\left[
1-f_{+}(\varepsilon_{k})
\right]
\right]
\label{theremodynamicpotential}
\end{align}
with 
$\varepsilon_{k}
=\sqrt{k^2+m^2_{\text{D}}}$
being the single-particle dispersion.
Here, 
$\text{tr}_{\text {D}}$
is the trace over the spinor indices, respectively.
In the second line, we introduced the Fermi-Dirac distribution function
$f_{\pm}(z)$
\begin{align}
f_{\pm}(z)=
\frac{1}{e^{\beta (z \pm \mu^{*})}+1}.
\end{align}
The antiparticle contribution is retained here for completeness, although it will be exponentially suppressed in the low-temperature dense-matter regime considered later.

To characterize the quasiparticle excitations near the Fermi surface within the mean-field approximation, we introduced the quasiparticle Green’s function
$G(i\omega_{n}, \bm{k})$
and extract the corresponding dispersion relation.
The inverse of the Green’s function is defined by
\begin{align}
G^{-1}(i\omega_{n}, \bm{k})
=
\gamma^{0}(i\omega_{n}-\mu^{*})
-\gamma^{i}k_{i}+m_{\text{D}}
\end{align}
with fermionic Matsubara frequency 
$\omega_{n}=(2n+1)\pi/\beta$
originating from the anti-periodicity of the nucleon field 
$\psi$.
The poles of the analytically continued Green's function determine the quasiparticle dispersion relation.
By taking the analytic continuation
$i\omega_{n} \to \xi+i0$,
we obtain
\begin{align}
(\xi+\mu^{*})^2-\varepsilon^2_{k}=0.
\end{align}
The positive-energy solution defines the quasiparticle excitation energy measured from the effective chemical potential
\begin{align}
\xi_{k}
=
\sqrt{k^2+m^2_{\text{D}}}-\mu^{*}
=
\varepsilon_{k}-\mu^{*}.
\end{align}
The Fermi surface is specified by the condition
$\xi_{k_{\text{F}}}=0$,
namely,
\begin{align}
\mu^{*}
=
\varepsilon_{k_{\text{F}}}
=
\sqrt{k^2_{\text{F}}+m^2_{\text{D}}}
\end{align}
with the Fermi momentum
$k_{\text{F}}$.
The Fermi velocity is therefore given by
\begin{align}
v_{\text{F}}
=
\left.
\frac{\partial \xi_{k}}{\partial k}
\right|_{k=k_{\text{F}}},
\end{align}
and the Landau mass
$m_{\text{L}}$
is defined by the ratio of the Fermi momentum to the Fermi velocity~\cite{Typel2005,Peter2022},
\begin{align}
m_{\text{L}}
=
\frac{k_{\text{F}}}{v_{\text{F}}}.
\end{align}
Using
$\varepsilon_{k}
=\sqrt{k^2+m^2_{\text{D}}}$,
we find
\begin{align}
v_{\text{F}}
=
\left.
\frac{\partial \xi_{k}}{\partial k}
\right|_{k=k_{\text{F}}}
=
\frac{k_{\text{F}}}{\sqrt{k^2_{\text{F}}+m^2_{\text{D}}}},
\end{align}
and hence
\begin{align}
m_{\text{L}}
=
\sqrt{k^2_{\text{F}}+m^2_{\text{D}}}.
\label{eq:mLD}
\end{align}
This relation shows that, within the RMF theory, 
the Landau mass is determined by the quasiparticle dispersion at the Fermi surface and is, in general, distinct from the Dirac mass
$m_{\text{D}}$.

The mean-field values 
$\sigma$ 
and 
$\omega$
should satisfy the following stationary conditions
\begin{align}
\frac{\partial \Omega}{\partial \sigma}=0,
\quad
\frac{\partial \Omega}{\partial \omega}=0,
\end{align}
which lead to the self-consistent equations
\begin{align}
m^2_{\sigma}\sigma
=
g_{\sigma}n_{\text{s}}
,
\quad
m^2_{\omega}\omega
=
g_{\omega}n,
\label{selfconsistent}
\end{align}
where the quantities
$n_{\text{s}}$
and
$n$
are the scalar density and the baryon density, respectively
\begin{align}
n_{\text{s}}
=
2
\int \frac{d^3k}{(2\pi)^3} 
\frac{m_{\text{D}}}{\varepsilon_{k}}
\left[
f_{-}
\left(
\varepsilon_{k}
\right)
+
f_{+}
\left(
\varepsilon_{k}
\right)
\right],
\quad
n
=
2
\int \frac{d^3k}{(2\pi)^3} 
\left[
f_{-}
\left(
\varepsilon_{k}
\right)
-
f_{+}
\left(
\varepsilon_{k}
\right)
\right].
\end{align}
These quantities suggest that the number of scalar and baryon are related to the occupied number of the nucleon.
Note that, since the scalar density 
$n_{\text{s}}$
depends on the Dirac mass
$m_{\text{D}}$,
the latter is determined self-consistently through
$m_{\text{D}}=m_{\psi}-g_{\sigma}\sigma
=m_{\psi}-g^2_{\sigma}n_{\text{s}}/m^2_{\sigma}$.
Thus, the Dirac mass
$m_{\text{D}}$
represents the in-medium nucleon mass term reduced from the free nucleon mass 
$m_{\psi}$
by the attractive scalar self-energy.

We next derive the specific heat per unit volume
$c_{\text{v}}$.
For this purpose, we evaluate the energy density 
$\epsilon$
and the entropy density 
$s$ 
while keeping the baryon density 
$n$
fixed.
In case of the equilibrium conditions being satisfied, the entropy density is defined by the thermodynamic potential as 
\begin{align}
s=
-\left(
\frac{\partial \Omega}{\partial T}
\right)_{\mu, \sigma, \omega}.
\end{align}
A straightforward calculation gives
\begin{align}
s
&=
-2
\int \frac{d^3k}{(2\pi)^3}
\left[
\log
\left[
1-f_{-}(\varepsilon_{k})
\right]
+
\log
\left[
1-f_{+}(\varepsilon_{k})
\right]
\right]
-\beta \mu^{*} n
+2\beta
\int \frac{d^3k}{(2\pi)^3}
\varepsilon_{k}
\left[
f_{-}(\varepsilon_{k})+f_{+}(\varepsilon_{k})
\right].
\end{align}
Then, the energy density 
$\epsilon=\Omega + Ts + \mu n$ 
is given by
\begin{align}
\epsilon
&=
\frac{1}{2}m^2_{\sigma}\sigma^2
+\frac{1}{2}m^2_{\omega}\omega^2
+2\int \frac{d^3k}{(2\pi)^3}
\varepsilon_{k}
\left[
f_{-}(\varepsilon_{k})+f_{+}(\varepsilon_{k})
\right].
\end{align}

The specific heat per unit volume is defined by fixing the baryon density 
$n$
and the volume 
$V$
\begin{align}
c_{\text{V}}
=
\left(
\frac{d \epsilon}{d T}
\right)_{n}.
\label{specificheat}
\end{align}
Having established the finite-temperature mean-field framework and the relevant thermodynamic quantities, we now investigate the low-temperature thermodynamics at fixed baryon density.
In particular, we derive the leading temperature dependence of the energy density, specific heat per unit volume, as well as that of the self-consistent scalar mean field, which determines the finite-temperature correction to the Dirac mass.

\subsection{Low temperature approximation}
We now evaluate the low-temperature behavior of $c_{\text{V}}$. 
In the NS regime of interest, the temperature is much smaller than the Fermi energy, 
$T/\varepsilon_{k_{\text{F}}}\ll 1$.
In this regime, antiparticle contributions are exponentially suppressed, and the thermodynamics is governed by quasiparticle excitations in the vicinity of the Fermi surface~\cite{Baldo2012, Friman2019, Alford2022}.
We use the low-temperature expansion of the Fermi-Dirac distribution
\begin{align}
\frac{1}
{e^{\beta (\varepsilon_{k}-\mu^{*})}+1}
=
\theta
\left(
\mu^{*}-\varepsilon_{k}
\right)
-\frac{\pi^2 T^2}{6}
\left.
\frac{d\delta(\omega)}{d\omega}
\right|_{\omega=\varepsilon_{k}-\mu^{*}}
+\mathcal{O}
\left(
T^4
\right).
\label{FDapprox}
\end{align}
Note that this approximation is valid when 
$T \ll \mu^{*}
=\varepsilon_{k_{\text{F}}}$
is satisfied.
Then, the baryon density 
$n$
is approximated as
\begin{align}
n
\approx
2\int \frac{d^3k}{(2\pi)^3} 
\theta
\left(
\mu^{*}-\varepsilon_{k}
\right)
=
\frac{k^3_{\text{F}}}{3\pi^2},
\end{align}
where the cutoff 
$k_{\text{F}}$
appeared when
$\mu^{*}=\varepsilon_{k}$, i.e., 
$k=k_{\text{F}}$
is satisfied.
Note that the anti-particle part is exponentially suppressed of the NS parameters, and thus, we neglected the term.
One of the term in the thermodynamic potential is computed as
\begin{align}
\frac{2}{\beta}
\int \frac{d^3k}{(2\pi)^3}
\log
\left(
1+e^{-\beta(\varepsilon_{k}-\mu^{*})}
\right)
&=
\frac{1}{3\pi^2}
\int_{0}^{\infty}dk 
\frac{k^4}{\varepsilon_{k}}
\frac{1}{e^{\beta(\varepsilon_{k}-\mu^{*})}+1}
\nonumber\\
\quad
&\approx
\frac{1}{3\pi^2}
\int_{0}^{\infty}dk 
\frac{k^4}{\varepsilon_{k}}
\left[
\theta
\left(
\mu^{*}-\varepsilon_{k}
\right)
-\frac{\pi^2 T^2}{6}
\left.\frac{d\delta(\omega)}{d\omega}\right|_{\omega=\varepsilon_{k}-\mu^{*}}
\right]
\nonumber\\
\quad
&=
\frac{1}{3\pi^2}
\int_{0}^{k_{\text{F}}}dk 
\frac{k^4}{\sqrt{k^2+m^2_{\text{D}}}}
+\frac{\pi^2 T^2}{6}
\left(
\frac{k_{\text{F}}m_{\text{L}}}{\pi^2}
\right),
\end{align}
where, in the second line, we used the low-temperature approximation of Eq.~\eqref{FDapprox}.
Thus, the thermodynamic potential is rewritten as
\begin{align}
\Omega
\approx
\frac{1}{2}m^2_{\sigma}\sigma^2
-\frac{1}{2}m^2_{\omega}\omega^2
-\frac{1}{3\pi^2}
\int_{0}^{k_{\text{F}}}dk 
\frac{k^4}{\sqrt{k^2+m^2_{\text{D}}}}
-\frac{\pi^2 T^2}{6}
\left(
\frac{k_{\text{F}}m_{\text{L}}}{\pi^2}
\right).
\end{align}
The energy density 
$\epsilon$
is also approximated as
\begin{align}
\epsilon
&\approx
\epsilon_{0}
+
\frac{\pi^2 T^2}{6}
D(\varepsilon_{k})
\label{enerdenapprox}
\end{align}
with the zero temperature energy density $\epsilon_{0}$
\begin{align}
\epsilon_{0}
&=
\frac{1}{2}m^2_{\sigma}\sigma^2
+\frac{1}{2}m^2_{\omega}\omega^2
+\frac{1}{\pi^2}
\int_{0}^{k_{\text{F}}} dk k^2
\sqrt{k^2+m^2_{\text{D}}}.
\end{align}
Here, the density of states $D(\varepsilon_{k})$
is defined by
\begin{align}
D(\varepsilon_{k})
=
\frac{\partial n}{\partial \mu^{*}}
=
2\int \frac{d^3k}{(2\pi)^3} 
\delta
\left(
\mu^{*}-\varepsilon_{k}
\right)
=\int dk \frac{k^2}{\pi^2}
\delta
\left(
\mu^{*}-\varepsilon_{k}
\right).
\end{align}

The density of states 
$D(\varepsilon_{k})$
is proportional to the Landau mass.
Near the Fermi surface, the dispersion is linearized as
$
\varepsilon_{k}-\mu^{*}
\approx
v_{\text{F}}(k-k_{\text{F}})$
with $v_{\text{F}}
=k_{\text{F}}/m_{\text{L}}$.
Thus, the density of states is computed as
\begin{align}
D(\varepsilon_{k})
=
\int dk \frac{k^2}{\pi^2}
\frac{1}{v_{\text{F}}}
\delta
\left(k_{\text{F}}-k\right)
=
\frac{k^2_{\text{F}}}{\pi^2 v_{\text{F}}}
=
\frac{3m_{\text{L}}n}{k^2_{\text{F}}}
\label{dos},
\end{align}
where, in the last equality, we used 
$m_{\text{L}}=k_{\text{F}}/v_{\text{F}}$
and
$n=k^3_{\text{F}}/3\pi^2$.
The last equality suggests that the Landau mass 
$m_{\text{L}}$
characterizes the density of quasiparticle states near the Fermi surface~\cite{Typel2005, Zhang2016, Friman2019, Chamseddine2023, Alp2025}.

The connection between
$c_{\text{V}}$
and the Landau mass follows from the density of states at the Fermi surface.
Using the definition of
$c_{\text{V}}$
in Eq.~\eqref{specificheat} together with the low-temperature expansion of the energy density in Eq.~\eqref{enerdenapprox}, and substituting the density of states in Eq.~\eqref{dos}, we obtain
\begin{align}
c_{\text{V}}
&=
\frac{\pi^2 T}{3}
D(\varepsilon_{k})
=
\frac{m_{\text{L}}n}{k^2_{\text{F}}} \pi^2 T.
\end{align}
This is the standard Fermi-liquid result,
$c_{\text{V}} \propto T$,
whose coefficient is determined by the Landau mass through the density of states at the Fermi surface~\cite{Page2004,Friman2019}.

To show analytical formula of the mean-field scalar sector with finite temperature correction, we further expand the scalar mean-field up to the second order as
\begin{align}
\Delta_{\sigma}(T)
&=
\Delta_{\sigma}(0)
+
T
\left(
\left.
\frac{d}{d T}\Delta_\sigma(T)
\right|_{T=0}
\right)
+
\frac{T^2}{2}
\left(
\left.
\frac{d^2}{d^2 T}\Delta_\sigma(T)
\right|_{T=0}
\right)
+\mathcal{O}(T^3)
\nonumber\\
\quad
&=
\Delta_{\sigma}(0)
+\frac{\pi^2}{2}C^2_{\sigma}(1-\Delta_{\sigma}(0))
\frac{m_{\psi}}{k_{\text{F}}}
\sqrt{
\left(
\frac{k_{\text{F}}}{m_{\psi}}
\right)^2
+\left(
1-\Delta_{\sigma}(0)
\right)^2
}
\left(
\frac{T}{m_{\psi}}
\right)^2
+\mathcal{O}(T^3),
\end{align}
where we introduced the new mean-field variable 
$\Delta_{\sigma}
=g_{\sigma}\sigma/m_{\psi}$
and the effective coupling constant
$C^2_{\sigma}
=(g^2_{\sigma}/3\pi^2)(m_{\psi}/m_{\sigma})^2$
defined in Ref.~\cite{Zurek2019}.
From the self-consistent equation Eq.~\eqref{selfconsistent} and the low-temperature approximation Eq.~\eqref{FDapprox}, we used 
$d \Delta_{\sigma}/ d T|_{T=0}=0$.
$\Delta_{\sigma}(0)$
satisfies the following self-consistent equation
\begin{align}
\Delta_{\sigma}(0)
=
3C^2_{\sigma}
\int_{0}^{k_{\text{F}}/m_{\psi}}dx x^2
\frac{
(1-\Delta_{\sigma}(0))
}{\sqrt{x^2+(1-\Delta_{\sigma}(0))^2}}.
\end{align}
Thus, the Dirac mass to order $\mathcal{O}(T^2)$ is given by
\begin{align}
m_{\text{D}}(T)
&=
m_{\psi}-g_{\sigma}\sigma(T)
\nonumber\\
\quad
&=
m_{\psi}(1-\Delta_{\sigma}(0))
\left[
1-\frac{\pi^2}{2}C^2_{\sigma}
\frac{m_{\psi}}{k_{\text{F}}}
\sqrt{
\left(
\frac{k_{\text{F}}}{m_{\psi}}
\right)^2
+\left(
1-\Delta_{\sigma}(0)
\right)^2
}
\left(
\frac{T}{m_{\psi}}
\right)^2
\right]
\label{eq:finiteT}
\end{align}
The leading finite-temperature correction is negative and proportional to 
$T^2$. 
Moreover, the coefficient of the negative 
$T^2$
correction becomes larger as 
$k_{\mathrm{F}}/m_{\psi}$
decreases. 
Consequently, finite-temperature effects lead to a stronger reduction of the Dirac mass at lower Fermi momenta, within the range of validity of the low-temperature expansion.
In the next subsection, we examine the resulting temperature and density dependence of the Dirac mass and compare it with that of the Landau mass.

\subsection{Behaviors of the Dirac mass and the Landau mass}
\begin{figure}[H]
 \centering
  \includegraphics[width=0.45\linewidth]{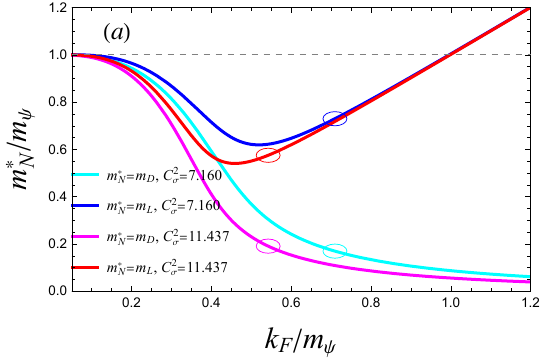}
  \includegraphics[width=0.45\linewidth]{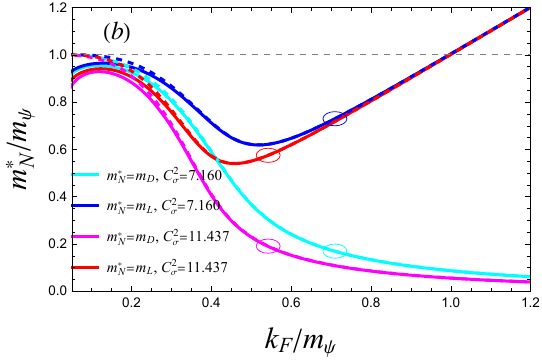}
   \caption
    {Behaviors of the Dirac mass and the Landau mass.
    We adopted the temperatures 
    $T/m_{\psi}=0$ for the panel (a).
    In the panel (b), we adopted
    $T/m_{\psi}=10/938$ (solid lines)
    and
    $T/m_{\psi}=0.01/938$ (dashed lines).
    }
\label{fig:mass}
\end{figure}
Fig.~\ref{fig:mass} shows the behaviors of the different effective nucleon masses
$m^{*}_{\text{N}}$
for the Dirac mass
$m^{*}_{\text{N}}=m_{\text{D}}$
and for the Landau mass
$m^{*}_{\text{N}}=m_{\text{L}}$
cases as a function of the normalized Fermi momentum
$k_{\text{F}}/m_{\psi}$.
Panel (a) corresponds to the zero-temperature case, while panel (b) shows the finite-temperature results for 
$T/m_{\psi}=10/938$ (solid lines)
and
$T/m_{\psi}=0.01/938$ (dashed lines).
Here, we choose the two values of the scalar coupling constant, 
$g_{\sigma}=7.93528$
and
$g_{\sigma}=10.0289$, 
listed in Table~1 of Ref.~\cite{2025ApJ...980...54L}.
These values correspond to 
$C_{\sigma}^2=7.160$ 
and
$C_{\sigma}^2=11.437$, respectively.
This choice allows us to compare the theoretical results from the Walecka model with the cooling results obtained using realistic EOS models in Sec.~III B.
For both
$C^2_{\sigma}=7.160$
and
$C^2_{\sigma}=11.437$, 
the Dirac mass decreases monotonically as the density increases, indicating a strong reduction of the effective nucleon mass in the high-density regime.
By contrast, the Landau mass exhibits a non-monotonic behavior: it first decreases, reaches a minimum at intermediate density, and then increases again at larger
$k_{\text{F}}/m_{\psi}$.
For the Dirac mass, with increasing baryon density
$n=k^3_{\text{F}}/3\pi^2$,
the Fermi momentum increases and a larger number of nucleon quasiparticle states are occupied. 
As a result, the scalar density 
$n_{\text{s}}$,
which provides the source for the scalar mean-field, becomes larger. 
The enhanced attractive scalar mean-field then lowers the in-medium mass term, leading to a reduction of the Dirac mass 
$m_{\text{D}}$.
For the Landau mass, the density dependence should be interpreted through the quasiparticle dispersion near the Fermi surface. 
The scalar mean-field modifies the in-medium nucleon mass term, while the Landau mass is determined by the Fermi momentum and the Fermi velocity, 
$m_{\text{L}}=k_{\text{F}}/v_{\text{F}}$.
Thus, although the Dirac mass 
$m_{\text{D}}$
decreases due to the attractive scalar mean-field, the Landau mass can increase at high densities because of the growth of the Fermi momentum 
$k_{\text{F}}$.
As a result, the difference between the Dirac mass and the Landau mass becomes increasingly pronounced in the high-density region. 

Once the relation between the energy density and pressure is obtained, the NS structure (e.g., mass-radius relation or mass-central density relation) can be determined by solving Tolman-Oppenheimer-Volkoff (TOV) equations~\cite{1939PhRv...55..364T,1939PhRv...55..374O}. 
Since the EOS specifies how much pressure is available to support matter against gravity, it uniquely determines the maximum central density attainable in a stable NS configuration. 
To see how effective mass changes at most, we extract the maximum central 
$k_{\text{F}}/m_{\psi}$
values from maximum-mass NS for each EOS model, which are plotted as the circles in Fig.~\ref{fig:mass}. 
Compared to the Laudau mass at the maximum mass density, the corresponding Dirac mass becomes lower by a factor of 
$\sim 4 (3)$
for 
$C^2_{\sigma}=7.160 (11.437)$.
As we discuss in the next section, this difference between the two effective masses has a clear impact on the cooling behavior.

Regarding the finite-temperature effect, one can see the difference mainly at low density, whereas it becomes negligible at high density. 
This is because the relevant thermal scale is small compared with the Fermi energy in the dense regime, so thermal smearing around the Fermi surface is strongly suppressed. 

Thus, in the RMF framework, the Landau mass
$m_{\text{L}}$
should be distinguished from the Dirac mass
$m_{\text{D}}$.
While the Dirac mass represents the in-medium mass term modified by the scalar mean-field, the Landau mass characterizes the quasiparticle density of states at the Fermi surface.
To quantitatively analyze the difference of the Dirac and Landau masses, we investigate the behavior of the cooling curves for the neutron-neutron bremsstrahlung and direct Urca process in the next section.

\section{Cooling curves}
The thermal evolution of a NS is governed by the competition between its heat capacity and neutrino energy losses. 
Since both quantities depend sensitively on the quasiparticle properties near the Fermi surface, the choice of the effective mass can have a sizable impact on the cooling curve.
In particular, the Dirac mass and the Landau mass enter different physical quantities: the former characterizes the scalar mean-field contribution to the single-particle spectrum, whereas the latter controls the density of states and hence the low-temperature specific heat.

In this section, we investigate how different prescriptions for the nucleon effective mass affect the cooling behavior. 
We consider two representative neutrino-emission mechanisms.
The first is neutron-neutron bremsstrahlung, which provides a slow cooling channel in non-superfluid neutron matter. 
The second is the nucleon direct Urca process, which gives a fast cooling channel in charge-neutral \(npe\) matter when the momentum-conservation condition at the Fermi surfaces is satisfied.

Generally, temperature is density dependent due to the difference in the locally neutrino cooling rate, which results in a colder core compared to the crust. 
However, as most of the observed NSs are older than the thermal relaxation timescale 
$\sim10$--$100~{\rm yr}$~\citep{1994ApJ...425..802L,2001MNRAS.324..725G}, 
we here describe the thermal evolution assuming the isothermal cooling equation as
\begin{align}
c_{\text{V}}\frac{dT}{dt}
=-\mathcal{E}_{\nu},
\end{align}
where 
$\mathcal{E}_{\nu}$
denotes the total neutrino emissivity, namely the energy-loss rate per unit volume due to neutrino emission. 
In the following, we consider two representative choices of 
$\mathcal{E}_{\nu}$: 
neutron-neutron bremsstrahlung for the slow cooling channel
($\mathcal{E}_{\nu}
=\mathcal{E}_{\text{NN}}$)
and the nucleon direct Urca process for the fast cooling channel
($\mathcal{E}_{\nu}
=\mathcal{E}_{\text{DU}}$).

\subsection{Theoretical cooling calculation}
We first consider the slow cooling channel in non-superfluid neutron matter. 
In this case, the dominant neutrino-emission process is neutron-neutron bremsstrahlung,
$n+n\rightarrow n+n+\nu+\bar{\nu}$.
In the non-relativistic regime, the corresponding emissivity can be written as~\cite{Friman1979, Yakovlev2001}
\begin{align}
\mathcal{E}_{\text{NN}}
=
\frac{41 N_{\nu}}{14175}
\frac{G^2_{\text{F}}C^2_{\text{A}}m^{*4}_{\text{N}}}{2\pi}
\left(
\frac{f_{\pi \text{NN}}}{m_{\pi}}
\right)^4
p_{\text{n}}F\left(\frac{m_{\pi}}{2p_{\text{n}}}\right)T^8
\end{align}
where $N_\nu$ denotes the number of flavor and a function $F(x)$ satisfying
\begin{align}
F(x)=
1-\frac{3}{2}x\tan^{-1}\left(\frac{1}{x}\right)
+\frac{1}{2}\left(\frac{x^2}{1+x^2}\right).
\end{align}
In the numerical comparison below, we evaluate this expression using either $m^{*}_{\text{N}}=m_{\text{D}}$
or
$m^{*}_{\text{N}}=m_{\text{L}}$, 
in order to isolate the impact of the effective-mass prescription.

For the bremsstrahlung cooling channel, the local cooling equation becomes
\begin{align}
c_{\text{V}}\frac{dT}{dt}
=-\mathcal{E}_{\text{NN}}.
\end{align}
We solve this equation for a homogeneous isothermal core, using the low-temperature specific heat derived in the previous section.
\begin{figure}[H]
 \centering
  \includegraphics[width=0.47\linewidth]{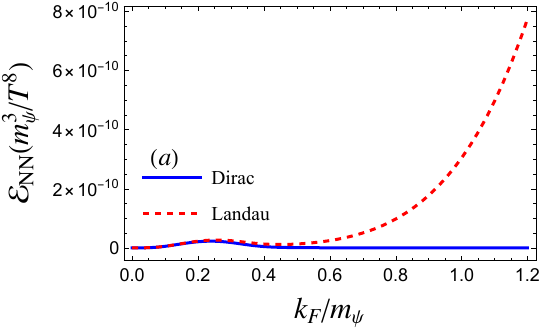}
  \includegraphics[width=0.47\linewidth]{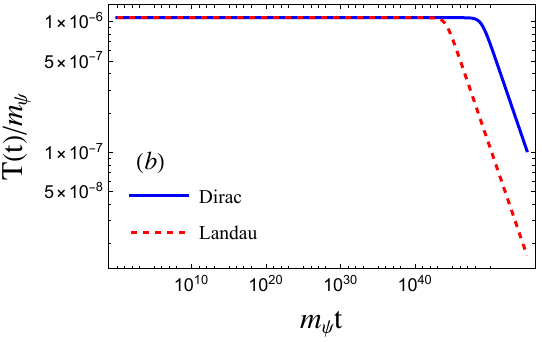}
   \caption
    {Slow cooling example based on neutron-neutron bremsstrahlung.
    (a) Neutron-neutron bremsstrahlung emissivity divided by the temperature
    $T^8$
    as a function of 
    $k_{\text{F}}/m_{\psi}$.
    We fixed $N_{\nu}=2$.
    (b) Temperature evolution obtained from the cooling equation at fixed 
    $k_{\text{F}}/m_{\psi}=1$.
    The solid blue and dashed red curves correspond to the Dirac mass and Landau mass prescriptions, respectively.
    }
\label{fig:emissivitiesNN}
\end{figure}
Figure~\ref{fig:emissivitiesNN} shows the effect of the effective mass prescription on the bremsstrahlung emissivity and on the resulting cooling curve. 
Panel (a) denotes 
$\mathcal{E}_{NN}$
normalized by
$m^3_{\psi}/T^8$ at fixed 
$k_F/m_\psi=1$, 
while panel (b) shows the temperature evolution
as a function of 
$k_F/m_\psi$. 
Since the bremsstrahlung emissivity contains a high power of the effective mass, the difference between the Dirac mass and Landau mass prescriptions leads to a significant change in the cooling rate. 
In particular, using the Landau mass enhances the emissivity in the high-density region and results in a slower decrease of the temperature in the parameter range shown.

We next consider a fast cooling channel in charge-neutral \(npe\) matter. 
For the fast cooling case, the cooling equation is given by
\begin{align}
c_{\text{V}}\frac{dT}{dt}
&=-\mathcal{E}_{\text{DU}}
\end{align}
with, in the relativistic regime, 
$\mathcal{E}_{\text{DU}}
=\mathcal{E}^{\text{R}}_{\text{DU}}$
and, in the non-relativistic regime, 
$\mathcal{E}_{\text{DU}}
=\mathcal{E}^{\text{NR}}_{\text{DU}}$
being satisfied.
When the proton fraction is sufficiently large, the nucleon direct Urca process,
$n \rightarrow p+e^{-}+\bar{\nu}_{\text{e}}$,
and
$p+e^{-} \rightarrow n+\nu_{\text{e}}$,
is allowed by momentum conservation at the Fermi surfaces. 
The kinematic condition is
\begin{align}
p_{\text{p}}+p_{\text{e}} \ge p_{\text{n}}
\end{align}
with the Fermi momentum of the proton, electron, and neutron,
$p_{\text{p}}$, 
$p_{\text{e}}$, 
and
$p_{\text{n}}$,
respectively.
For degenerate matter, the emissivity up to the lowest order in 
$\mu_{\text{e}}/T$
in the relativistic treatment is given by~\cite{Leinson2001, Leinson2002}
\begin{align}
\mathcal{E}^{\text{R}}_{\text{DU}}
=
\frac{457\pi}{10080}
G^{2}_{\text{F}}C^2T^6
&
\Big[
C_{\text{V}}C_{\text{A}}
\left(
(\epsilon_{\text{n}}+\epsilon_{\text{p}})
p^2_{\text{e}}
-(\epsilon_{\text{n}}-\epsilon_{\text{p}})
(p^2_{\text{n}}-p^2_{\text{p}})
\right)
+
(C^2_{\text{A}}-C^2_{\text{V}})
\mu_{\text{e}}m^{*}_{\text{N}}m^{*}_{\text{p}}
\nonumber\\
\quad
&
+(C^2_{\text{V}}+C^2_{\text{A}})
\left(
2\mu_{\text{e}}\epsilon_{\text{n}}\epsilon_{\text{p}}
+\epsilon_{\text{n}}p^2_{\text{e}}
-\frac{1}{2}(\epsilon_{\text{n}}+\epsilon_{\text{p}})
(p^2_{\text{n}}+p^2_{\text{e}}-p^2_{\text{p}})
\right)
\Big]
\theta(p_{\text{p}}+p_{\text{e}}-p_{\text{n}}).
\label{eq:durel0}
\end{align}
Here, the quantities
$\epsilon_{\text{n}}
=\sqrt{p^2_{\text{n}}+m^2_{\text{D}}}$,
$\epsilon_{\text{p}}
=\sqrt{p^2_{\text{p}}+m^2_{\text{D}}}$
are the neutron and proton Fermi energies, respectively.
When 
$m^{*}_{\text{N}}$
and 
$m^{*}_{\text{p}}$
in the direct-Urca emissivity are identified with the Landau masses, they are given by
$m^{*}_{\text{p}}
=\sqrt{p^2_{\text{p}}+m^2_{\text{D}}}$
and
$m^{*}_{\text{N}}
=\sqrt{p^2_{\text{n}}+m^2_{\text{D}}}$.
Alternatively, when the effective nucleon masses entering the emissivity are identified with the Dirac masses, both the neutron and proton masses are given by
$m^{*}_{\text{N}}=m_{\text{D}}$
and
$m^{*}_{\text{p}}
=m_{\psi}-g_{\sigma}\sigma
=m_{\text{D}}$.
Here, we neglect the small difference between the neutron and proton masses and use a common bare nucleon mass 
$m_{\psi}$. 
Moreover, because the present model contains no isovector scalar mean field, the neutron and proton acquire the same in-medium Dirac mass.
The charged lepton is taken to be the electron, and its chemical potential is
$\mu_{\text{e}}
=\sqrt{p^2_{\text{e}}+m^2_{\text{e}}}$
with the electron mass 
$m_{\text{e}}$.
The Fermi momenta are related to the number densities by
$n_{i}={p_i^3}/{3\pi^2}$
with
$i=\text{n}, \text{p}, \text{e}$.
Charge neutrality implies
$n_{\text{e}}=n_{\text{p}}$
and therefore
$p_{\text{e}}=p_{\text{p}}$.
Introducing the baryon density 
$n_{\text{B}}
=n_{\text{{n}}}+n_{\text{{p}}}$,
the electron Fermi momentum
$p_{\text{e}}$
is related to the electron fraction
$Y_{\text{e}}=n_{\text{e}}/n_{\text{B}}$
as
$p_{\text{e}}
=(3\pi^2 Y_{\text{e}} n_{\text{B}})^{1/3}$.
Equivalently, the proton and electron Fermi momenta can be written in terms of the neutron Fermi momentum as
$p_{\text{p}}
=
p_{\text{e}}
=
p_{\text{n}}
\left(
Y_{\text{p}}/Y_{\text{n}}
\right)^{1/3}$
with the neutron and the proton fractions
$Y_{\text{n}}
=n_{\text{n}}/n_{\text{B}}$
and
$Y_{\text{p}}
=n_{\text{p}}/n_{\text{B}}$, respectively.
For simplicity, we impose charge neutrality and assume constant particle fractions,
$Y_{\text{e}}=Y_{\text{p}}=0.2$
and 
$Y_{\text{n}}=1-Y_{\text{p}}=0.8$.

In the non-relativistic limit for nucleons, 
$p_{\text{n}}, p_{\text{p}} 
\ll m^{*}_{\text{N}}$~\cite{Leinson2001, Leinson2002}, 
while the electrons are kept relativistic, the direct Urca emissivity reduces to
\begin{align}
\mathcal{E}^{\text{NR}}_{\text{DU}}
=
\frac{457\pi}{10080}
G^{2}_{\text{F}}C^2T^6
(C^2_{\text{V}}+3C^2_{\text{A}})
m^{*}_{\text{N}}m^{*}_{\text{p}}\mu_{\text{e}}
\theta(p_{\text{p}}+p_{\text{e}}-p_{\text{n}}),
\end{align}
where the nucleons are assumed to be non-relativistic, while the electron chemical potential is kept as 
$\mu_{\text{e}}
=\sqrt{p_{\text{e}}^2+m^2_{\text{e}}}$.
\begin{figure}[H]
 \centering
  \includegraphics[width=0.47\linewidth]{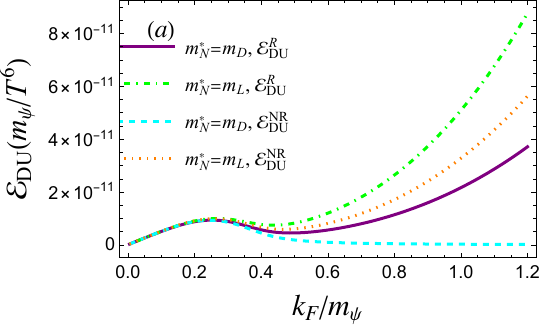}
  \includegraphics[width=0.47\linewidth]{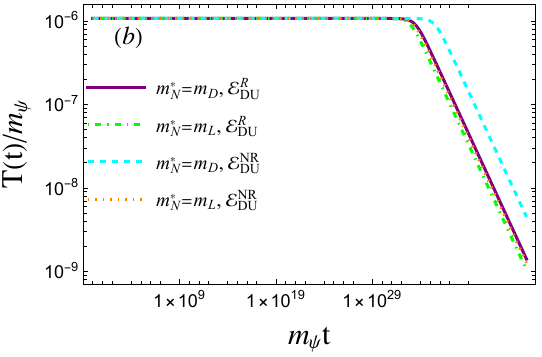}
   \caption
    {Fast cooling example based on the nucleon direct Urca process.
    (a) Direct-Urca emissivity divided by the temperature
    $T^6$ 
    as a function of 
    $k_{\text{F}}/m_{\psi}$.
    (b) Temperature evolution obtained from the cooling equation.
    The curves compare the relativistic emissivity 
    $\mathcal{E}^{\text{R}}_{\text{DU}}$ 
    and the non-relativistic approximation 
    $\mathcal{E}^{\text{NR}}_{\text{DU}}$, 
    evaluated with either 
    $m^{*}_{\text{N}}=m_{\text{D}}$ 
    or
    $m^{*}_{\text{N}}=m_{\text{L}}$.
    The particle fractions are fixed to 
    $Y_{\text{p}}=Y_{\text{e}}=0.2$ 
    and
    $Y_{\text{n}}=0.8$.
    }
\label{fig:DU}
\end{figure}
Figure~\ref{fig:DU} compares the direct-Urca emissivities and the corresponding cooling curves for the different effective mass prescriptions. 
Panel (a) shows 
$\mathcal{E}_{\text{DU}}$
normalized by
$m_{\psi}/T^6$
as a function of 
$k_{\text{F}}/m_{\psi}$, 
while panel (b) shows the temperature evolution obtained from the cooling equation at 
$k_{\text{F}}/m_{\psi}=1$.
We show both the relativistic expression 
$\mathcal{E}^{\text{R}}_{\text{DU}}$
and the non-relativistic approximation 
$\mathcal{E}^{\text{NR}}_{\text{DU}}$. 
In these plots, the proton and electron fractions are kept fixed, 
$Y_{\text{p}}=Y_{\text{e}}=0.2$, 
so that the direct-Urca threshold condition is satisfied in the range shown. 
The comparison illustrates that the cooling curve is sensitive not only to the neutrino emission channel but also to the effective mass prescription used in the emissivity and heat capacity.
The parameters to draw the above figures are summarized in the following table~\ref{tab:parameter}.
Note that we use the dimensionless Fermi coupling constant
$G_{\text{F}} m^2_{\psi}$
where
$G_{\text{F}}=
1.166 \times 10^{-11} \mathrm{~[MeV^{-2}]}$
is the conventional Fermi coupling constant.
\begin{table}[H]
  \centering
  \begin{tabular}{|c|c|c|c|}
   \hline
   Symbol & Description & Value \\
   \hline
$C_{\text{V}}$ & vector coupling & $1.0$ \\
$C_{\text{A}}$ & axial-vector coupling & $1.26$ \\
$C$ & Cabibbo factor & $0.973$ \\
$f_{\pi \text{NN}} $ & pion-nucleon coupling constant & $ 1.0$ \\
\hline
$G_{\text{F}}m^2_{\psi} $ & (dimensionless) Fermi coupling constant & $1.026 \times 10^{-5}$ \\
$m_{\text{e}} \mathrm{~[MeV]}$ & electron mass & $0.511$ \\
$m_{\psi} \mathrm{~[MeV]}$ & neutron mass & $938.0$ \\
$m_{\pi} \mathrm{~[MeV]}$ & pion mass & $ 135.0$ \\ 
\hline
  \end{tabular}
    \caption{Parameters}
    \label{tab:parameter}
\end{table}
The analysis in this subsection clarifies how the Dirac and Landau masses enter the cooling equation through the heat capacity and neutrino emissivities. 
Although the calculation is simplified, it shows that the choice of effective mass prescription can modify the cooling efficiency even within the same neutrino emission channel. 
In the next subsection, we include these ingredients in a more realistic cooling setup and examine how the distinction between the two masses is reflected in the cooling curves.

\subsection{Realistic cooling calculation}
The previous section is based on somehow theoretical treatment, but the realistic cooling curves are rather complicated due to the inclusion of many model parameters. 
To verify the impact of effective masses, we perform cooling calculation with use of public code \texttt{NSCool}~\cite{2016ascl.soft09009P}, 
which solves the general relativistic heat diffusion equation under 
$T=0$
TOV profile given an EOS. 

\begin{figure}[t]
    \centering
    \begin{minipage}{0.322\linewidth}
    \includegraphics[width=\linewidth]{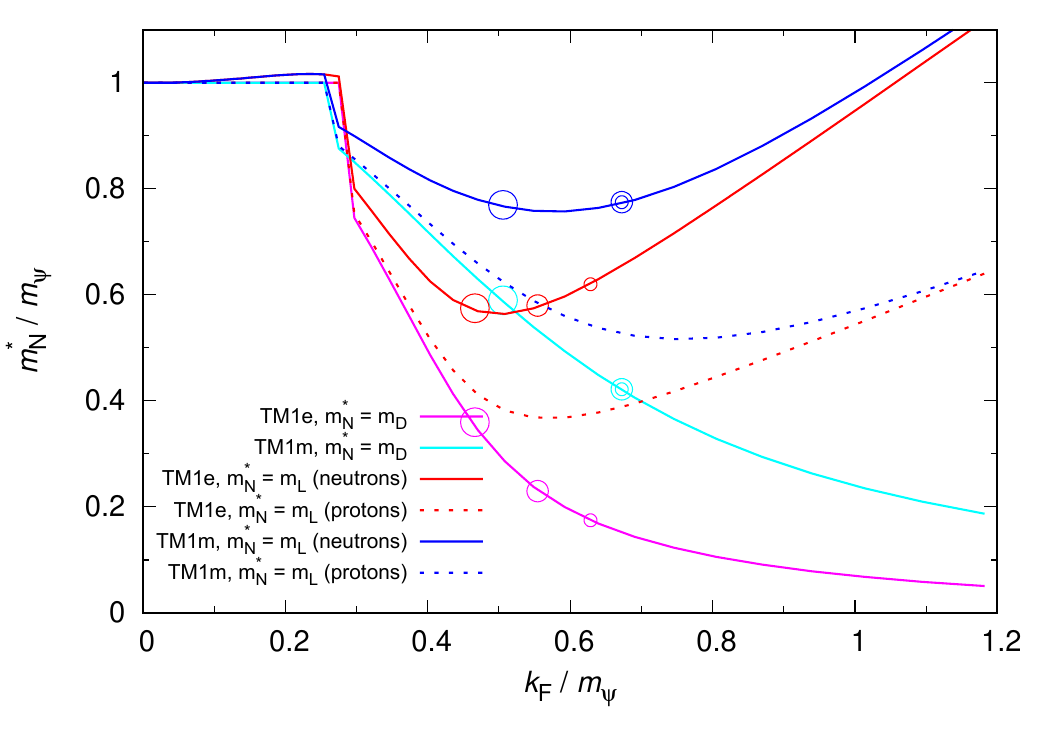}    
    \end{minipage}
        \begin{minipage}{0.322\linewidth}
\includegraphics[width=\linewidth]{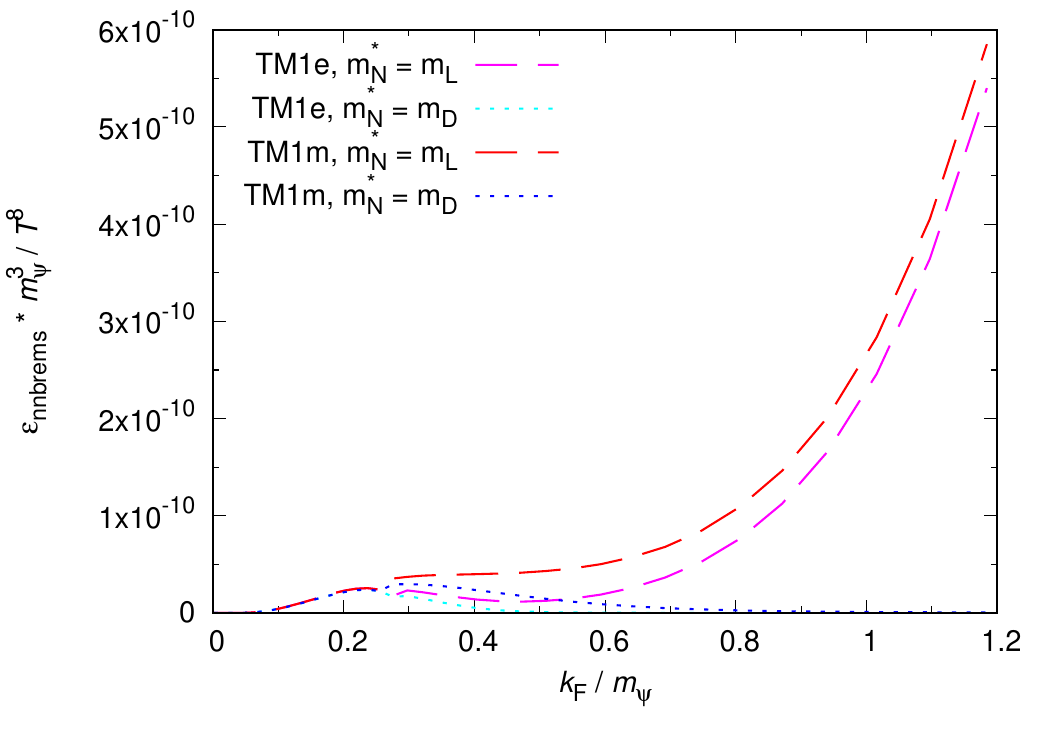}
    \end{minipage}
    \begin{minipage}{0.322\linewidth}
\includegraphics[width=\linewidth]{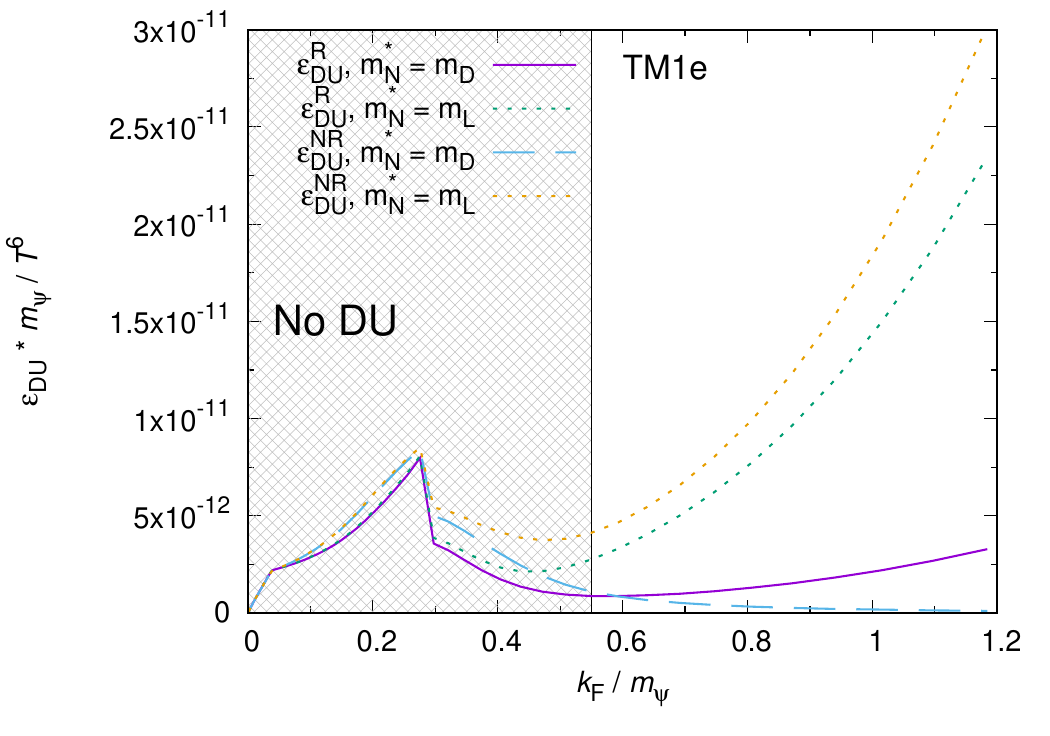}    
    \end{minipage}
    \caption{Baryon Fermi-wave number density dependence of some quantities:\\
    Left panel: Effective mass ratios with TM1e and TM1m EOS models. 
    The large, middle, and small circles indicate the central value of $k_{\text{F}}/m_{\psi}$ with $M/M_{\odot}=$1.4, 2.0, and 2.1. 
    Middle panel: Emissivities of the neutron-neutron bremsstrahlung with the TM1e and TM1m EOS. 
    Right panel: Emissivity of the DU process with the TM1e. 
    The gray hatched regions correspond to the density regions where the DU process is kinematically forbidden. 
    Hence, the emissivity below the DU threshold density (drawn as the black vertical line) is actually zero, although we plot in those density regions by replacing 
    $\theta(p_{\text{p}}+p_{\text{e}}-p_{\text{n}})$ 
    in Eq. (\ref{eq:durel0}) into 1.
    }
    \label{fig:edu_tm1e}
\end{figure}

Although thermodynamic amounts on the EOS were calculated with 
$\sigma$-$\omega$ 
model in Section II, we assumed a single nucleon with a free mass 
$M_{\text{N}}$
as the internal composition, which is, needless to say, essential for cooling curves. 
As more sophisticated nuclear models, we employ two EOS of 
TM1m~\cite{2025ApJ...980...54L}
and 
TM1e~\cite{2020ApJ...891..148S}, 
both of which are based on 
$\sigma\omega\rho$ 
RMF theory. 
TM1m was recently constructed by replacing the RMF parameter sets with those of TM1e, solely to adjust the effective mass while maintaining the saturation properties of nuclear matter. 
The left panel of Fig.~\ref{fig:edu_tm1e} displays the density dependence of effective mass ratios. 
The new TM1m model shows higher 
$m_{\text{D}}$
than TM1e due to the choice of lower effective mass at the saturation density (see Table 2 in \cite{2025ApJ...980...54L}). In both EOS models, neutron $m_{\text{L}}$
becomes high compared to 
$m_{\text{D}}$, 
especially in high-density regions, which is in good agreement with our theoretical models with 
$C^2_{\sigma}=7.160$ 
and 
$C^2_{\sigma}=11.437$ 
in Figs.~\ref{fig:mass}. Due to the neutron-rich environment, proton $m_{\text{L}}$ is lower than the neutron's one, although the density dependence is quite similar.
The maximum central density is similar between the Walecka model and 
$\sigma\omega\rho$ models, 
although there is a small discrepancy due to the absence of 
$\rho$ and $\omega\rho$
mediated interaction to nucleon in the former.

While the new TM1m model prohibited the DU process, TM1e allows it at the critical mass 
$M_{\text{DU}} = 2.06~M_{\odot}$. 
Fig.~\ref{fig:edu_tm1e} presents the density dependence of the emissivity with the TM1e. 
As already found by~\cite{Leinson2001}, the non-relativistic DU process with 
$m^{*}_{\text{N}}=m_{\text{D}}$
decreases with the density, while the relativistic one increases. 
If we take 
$m^{*}_{\text{N}}=m_{\text{L}}$, 
the fully non-relativistic DU process is close to (a bit weaker than) the fully relativistic DU process. 
In that sense, the relativistic effects on the DU process are not so significant. 
Regarding the neutron-neutron bremsstrahlung, while the emissivity increases with density in the case of Landau mass, it decreases with density in the case of Dirac mass. 
These features are in perfect agreement with the analysis in the Walecka model.

On the emissivity of the DU process, one may wonder about the difference by a factor of $\sim2$ in high-density regions, between the Walecka model (left panel in FIG. \ref{fig:DU}) and TM1e (right panel in FIG. \ref{fig:edu_tm1e}) despite similar density dependence of effective masses. 
This is attributed to the fact that TM1e has a small $Y_p$ value below 0.2 in all density regions (see also Fig. 4 in \cite{2022IJMPE..3150006D}). 
In fact, we also calculate the DU emissivity with one of the proton-rich EOSs, TM1 \cite{1998NuPhA.637..435S,1998PThPh.100.1013S}, 
which is the original version compared to TM1e and TM1m, and show that the DU emissivity is higher by 
$\sim20\%$
than FIG.~\ref{fig:DU} and similar to FIG.~1 in \cite{Leinson2001}. 
Hence, density dependence of $Y_p$ is indeed relevant not only with the threshold condition but also with the strength.

\begin{figure}[t]
    \centering
    \includegraphics[width=0.9\linewidth]{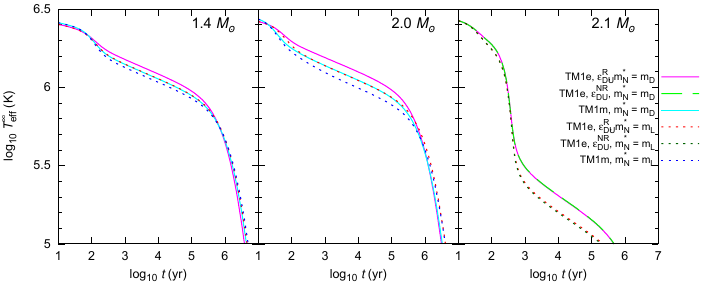}
    \caption{
    Cooling curves with different treatments of EOS models, effective masses, and emissivity of the DU process. 
    Note that we did not show cooling curves for 
    $M=2.1~M_{\odot}$
    with the TM1m EOS because the maximum mass is exceeded.
    }
    \label{fig:cc_eos}
\end{figure}

Fig.~\ref{fig:cc_eos} shows the impact of relativistic effects on the cooling curves of isolated NSs, i.e., age--effective temperature at infinite observer distance 
($T^\infty_{\rm eff}$) 
relations. 
In a slow cooling scenario with 
$M<2~M_{\odot}$, 
the surface temperature becomes lower with TM1m than with TM1e. 
Moreover, the difference of cooling curves with two EOS models becomes larger with 
$2~M_{\odot}$
stars than with 
$1.4~M_{\odot}$.
These are attributed to a higher effective mass with TM1m, which makes the modified Urca process and nucleon bremsstrahlung stronger by a factor of 
$(m^*_{\text{N}}/m_{\text{N}})^4$. 
The ratio of this factor between TM1m and TM1e is $7.2$, $11.3$, and $33.8$ with 
$1.4, 2.0$
and 
$2.1~M_{\odot}$ NSs, respectively (see also left panel in Fig.~\ref{fig:edu_tm1e}), which could increase the cooling rate with the same factor.

For massive NSs, the central 
$Y_{\text{p}}$
value tends to increase due to the symmetry energy in high-density matter, and finally, the DU process could be allowed. 
In TM1e EOS, this is realized at 
$2.06~M_{\odot}$, 
above which the NS cools rapidly as shown in the right panel in Fig.~\ref{fig:cc_eos}.
We found that the effect of relativity on the DU process is negligible with TM1e. 
As shown in Fig.~\ref{fig:edu_tm1e}, the difference between 
$\epsilon_{\text{DU}}^{\text{R}}$ 
and 
$\epsilon_{\text{DU}}^{\text{NR}}$ 
is small at the center of 
2.1 $M_{\odot}$
($k_{\text{F}}/m_{\psi}=0.63$) 
\footnote{
We also calculate the cooling curves with the TM1 EOS, which has the same effective masses as TM1e, but higher 
$Y_{\text{p}}$ 
due to a large slope of symmetry energy ($L$).
Then, the effect of relativity on the DU process shows faster cooling after reaching the thermal relaxation time. 
Thus, this feature depends on the EOS, in particular for the density dependence of $Y_{\text{p}}$.
}.
If a softer EOS near the central density is chosen, due to the larger difference between 
$\epsilon_{\text{DU}}^{\text{R}}$ 
and
$\epsilon_{\text{DU}}^{\text{NR}}$, 
cooling curves would also differ, but in this case, the threshold mass of the DU process becomes higher, and possibly, it may be prohibited. 
Also, the maximum mass generally decreases with softer EOS, which cannot support the existing 
$\sim2.1~M_{\odot}$
NSs~\cite{2020NatAs...4...72C,2021ApJ...918L..28M}. 
In fact, the softer EOS of TM1m has a maximum mass of 
$\simeq 2.0 M_{\odot}$, 
where the DU process does not operate. 
Overall, the relativity on the DU process has little impact on cooling curves. 

The impact of the Landau and Dirac masses is clearly reflected in the cooling curves.
In the slow-cooling scenario, the resulting variations in the cooling curves are comparable to those obtained with the TM1e and TM1m parameter sets. 
The corresponding differences in the surface temperature, arising from uncertainties in the effective masses and in the attractive scalar interaction, reach approximately 0.03(0.06) for 1.4(2.0) $M_\odot$ NSs.
These values are comparable to, and can even exceed, the observational uncertainties for well-studied cooling NSs, such as Cassiopeia A, whose typical uncertainty is of order 
$\sim 0.01$; see, e.g., \cite{2018ApJ...864..135P}.
Therefore, uncertainties in the effective masses constitute a non-negligible source of systematic uncertainty and should be taken into account when interpreting neutron-star cooling observations and constraining the properties of dense matter.
    
In the fast cooling case 
($2.1~M_\odot$
with the TM1e parameter set), NSs cool rapidly with the Landau mass compared to the Dirac one, similar to the slow cooling case.
However, the difference becomes visible only after the timescale of the most rapid cooling.
Before that timescale, because of a large temperature gradient between crust and core, the dominant neutrino cooling channel is an electron-ion bremsstrahlung, which does not depend on effective masses due to no relevancy with nucleon. 
Only cold middle-aged neutron stars with 
$T^\infty_{\rm eff}
\lesssim10^{5.5}~{\rm K}$
could be relevant with uncertainties of effective masses in the fast cooling scenario, although no such objects have been identified so far.

Cooling curves involve several uncertainties of conventional model parameters, such as nucleon superfluidity and the envelope composition, which have not been explored in the present study.
Indeed, the impact of these uncertainties on the cooling evolution may exceed that associated with the effective masses, making it challenging to constrain the latter from current cooling observations alone. 
Nevertheless, this does not dilute the physical importance of the effective masses. 
Unlike conventional model parameters, effective masses are fundamental quantities characterizing in-medium nuclear interactions and are difficult to determine directly from either terrestrial nuclear experiments or astrophysical observations.
Therefore, we stress that quantifying their influence on neutron-star cooling remains important for assessing the robustness of cooling predictions.

\section{Conclusion}
NSs are among the most compact objects in the Universe, containing matter at densities inaccessible to laboratory experiments.
In this study, we investigated the roles of two effective masses in NS cooling, focusing on the difference of Dirac mass and Landau mass, and the strength of the attractive interaction.

We firstly employed the Walecka model and introduced the Dirac and Landau masses within the RMF framework at finite temperature.
The Dirac mass is determined by the scalar mean-field, whereas the Landau mass characterizes the density of states near the Fermi surface.
We then examined the behavior of the two masses as functions of the Fermi momentum.
At low densities, the Dirac and Landau masses show similar behavior. 
However, at higher densities, their behavior becomes significantly different.
In this regime, the Dirac mass decreases due to the enhanced attractive scalar mean-field, whereas the Landau mass increases because it is dominated by the growth of the Fermi momentum.

We also demonstrated that increasing the scalar coupling 
$C^2_{\sigma}$
reduces the effective masses due to the attractive interaction being strong.
The Landau mass is affected indirectly through the resulting change in the quasiparticle spectrum.
Finite-temperature effects appear mainly in the low-density regime, where the Fermi energy is relatively small and thermal excitations near the Fermi surface are more important.
In contrast, at high densities and low temperatures, the system is strongly Fermi-degenerate, and thermal effects on the effective masses are suppressed.

We then computed the emissivities of neutron-neutron bremsstrahlung and the direct Urca process in order to clarify how the two effective masses affect neutrino emission.
We found that the difference between the Dirac and Landau masses becomes particularly important at high densities, consistently with the density dependence of the effective masses.

To verify the above results obtained within Walecka models, we performed NS cooling simulations with two sophisticated RMF EOSs, TM1e and TM1m, whose difference comes from the properties of nucleon effective masses. 
In terms of effective masses, these EOSs can be mimicked with Walecka models by adjusting $C^2_\sigma$ value. 
Then, we again confirmed faster cooling with the Landau mass than with the Dirac one, which tends to appear in more massive NSs. 
Also, in the slow cooling scenario, we found that TM1m shows faster cooling than TM1e due to the difference in effective masses. 
Since TM1e has a stronger attractive scalar interaction ($C^2_\sigma=11.437$) than TM1m ($C^2_\sigma=7.160$), this result is broadly consistent with our analysis within the Walecka model.
Although cooling curves are strongly influenced by other physical ingredients, such as superfluidity and envelope composition, uncertainties in the effective mass can induce changes that exceed observational errors in massive NSs.
Moreover, among cooling-model parameters, the most difficult quantity to be determined from nuclear experiments is the effective masses. 
As conventional model parameters are tightly constrained, uncertainties of effective masses must increase in importance.

The results of the present study suggest that the treatment of effective nucleon masses can have a sizable impact on neutron-star cooling. 
Nevertheless, several issues remain to be investigated. 
In this work, we focused on the scalar-meson mean-field within the RMF framework and did not include possible effects of nucleon superfluidity. 
Since superfluid components and pairing correlations can also modify the quasiparticle spectrum and the Dirac mass~\cite{Garani2022}, it will be important to incorporate these effects in future cooling simulations.

It would also be interesting to go beyond the momentum-independent(Hartree) RMF approximation adopted in this work. 
One possible direction is to include momentum-dependent self-energies, which can modify the relation between the Dirac and Landau masses and may affect the density of states near the Fermi surface~\cite{Typel1999, Typel2005}.
Another possible direction is to employ a dynamical mean-field approach, in which meson fields are treated as dynamical variables and the energy dependence of the quasiparticle self-energy is taken into account~\cite{Velikanov2021}.
Such extensions would allow a more systematic investigation of temperature- and density-dependent effective masses and their influence on neutrino emissivities and cooling curves.

Before closing this paper, let us comment on which definition of Laundan mass and Dirac mass is likely to describe the cooling curves, i.e., specific heat and neutrino emissivity in cold neutron stars. 
Since most of the nucleons contributing to them come from the neighborhood of the Fermi sea, the Landau mass is likely to be used, although this is indeed equivalent to the Dirac mass due to its momentum independence in the Hartree approximation. 
This equivalence holds even in the finite-temperature case because Eq. (\ref{eq:finiteT}) is momentum independent, at least as far as the Sommerfeld approximation is valid\footnote{
The Fermi energy (around typical saturation neutron momentum $p_n\sim1.7~{\rm fm}^{-1}$) 
is roughly
$E_F\approx p_n^2/(2m_n)
=60~{\rm MeV}(p_n/1.7~{\rm fm}^{-1})^2$,
which is much higher than the temperature inside the neutron-star core.}. 
Beyond the Hartree level, such as in the case with the Fock term, the Dirac mass in Eq. (\ref{eq:mLD}) becomes momentum dependent~\cite{Typel2005,Typel2010}, and only the Landau mass seems insufficient to describe cooling curves. 
Nevertheless, unless the NS is very massive, the impact of Fock terms on NS EOS is shown to be minor in the RMF framework~\cite{2016PhRvC..94d5803Z,2018EPJA...54..133L}. 
Therefore, even beyond the Hartree level, we believe that the nucleon mass relevant to cooling curves mostly comes from the Landau mass.

\acknowledgments
We thank Hiroyuki Tajima, Takumi Muto, and Daisuke Inotani for useful discussions. Y.S. is supported by JSPS KAKENHI Grant No. 25KJ0065. A.D. is supported by JSPS KAKENHI Grant No. JP25K17403.

\begin{appendix}
\section{
Derivation of Eq.~\eqref{theremodynamicpotential}}
In this appendix, we derive the fermionic contribution to the thermodynamic potential in Eq.~\eqref{theremodynamicpotential}. 
We first work in a finite cubic box of volume
$V=L^3$
with periodic boundary conditions. 
The spatial momenta are then discretized as
$\bm{k}
=2\pi\bm{n}/L$ 
($\bm{n}=(n_{x},n_{y},n_{z})\in \mathbb{Z}^{3}$).
Then, we use the normalized plane waves
\begin{align}
\langle
\tau,\bm{x}|n,\bm{k}
\rangle
=
\frac{e^{-i\omega_{n}\tau+i\bm{k}\cdot\bm{x}}}{\sqrt{\beta V}},
\label{planewave}
\end{align}
where 
$\omega_{n}=(2n+1)\pi/\beta$
with
$n \in \mathbb{Z}$
are fermionic Matsubara frequencies.
The inner product of this basis satisfies
\begin{align}
\int_{0}^{\beta} d\tau \int_{\text{V}} d^3x
\langle \tau, \bm{x}|n, \bm{k}\rangle
\langle n^{\prime}, \bm{k}^{\prime}|\tau, \bm{x}\rangle
=
\frac{1}{\beta V}
\int_{0}^{\beta} d\tau \int_{\text{V}} d^3x
e^{-i(\omega_{n}-\omega_{n^{\prime}})\tau+i(\bm{k}-\bm{k}^{\prime})\cdot\bm{x}}
=
\delta_{n, n^{\prime}}\delta_{\bm{k}, \bm{k}^{\prime}},
\label{orthonormal}
\end{align}
and the completeness relation
\begin{align}
\sum_{n,\bm{k}}|n,\bm{k}\rangle\langle n,\bm{k}|
=
\mathbb{1}.
\label{complete}
\end{align}
The thermodynamic limit
$V \to \infty$
leads to the continuum version of the momentum integral:
\begin{align}
\sum_{k}
\to
V\int 
\frac{d^3k}{(2\pi)^3}.
\end{align}

We now evaluate the functional trace
\begin{align}
\text{Tr}\log\mathcal{D},
\quad
\mathcal{D}
=
\gamma^{0}(\partial_{\tau}-\mu^{*})
-i\gamma^{i}\partial_{i}+m_{\text{D}}
\nonumber.
\end{align}
Here, 
$\mathrm{Tr}$ 
denotes the trace over spacetime, Matsubara frequency, momentum, and spinor indices. 
In the coordinate representation, the functional trace is written as
\begin{align}
\mathrm{Tr}\log \mathcal{D}
=
\int_{0}^{\beta} d\tau
\int_{\text{V}} d^3x
\,
\mathrm{tr}_{\mathrm D}
\left
\langle \tau,\bm{x}
\left|
\log \mathcal{D}
\right|\tau,\bm{x}
\right\rangle,
\end{align}
where 
$\mathrm{tr}_{\mathrm D}$
denotes the trace over Dirac matrix indices only.
We then insert the complete set of Matsubara plane waves~\eqref{complete}, which gives
\begin{align}
\mathrm{Tr}\log \mathcal{D}
&=
\sum_{n,\bm{k}}
\sum_{n',\bm{k}'}
\int_0^\beta d\tau \int_V d^3x\,
\langle \tau,\bm{x}|n,\bm{k}\rangle
\mathrm{tr}_{\mathrm D}
\left[
\left\langle n,\bm{k}\left|
\log\mathcal{D}
\right|n',\bm{k}'\right\rangle
\right]
\langle n',\bm{k}'|\tau,\bm{x}\rangle.
\end{align}
Here, the quantity
\begin{align}
\left\langle n,\bm{k}\left|
\log\mathcal{D}
\right|n',\bm{k}'\right\rangle
\end{align}
should be regarded as a matrix with respect to the spinor indices, labeled by the Fourier modes 
$(n,\bm{k})$
and
$(n',\bm{k}')$.

To evaluate this matrix element, we regard 
$\log\mathcal{D}$
as a function of the differential operator 
$\mathcal{D}$. 
Formally, this can be understood by expanding the logarithm in powers of 
$\mathcal{D}$. 
Then each power of 
$\mathcal{D}$
acts successively on the plane wave (complex conjugate of~\eqref{planewave})
\begin{align}
\langle n',\bm{k}'|
\tau,\bm{x}\rangle
=
\frac{1}{\sqrt{\beta V}}
e^{i\omega_{n'}\tau-i\bm{k}'\cdot\bm{x}}.
\nonumber
\end{align}
Since this plane wave is an eigenfunction of the derivative operators,
\begin{align}
\partial_{\tau}
\langle n',\bm{k}'|
\tau,\bm{x}\rangle
=
i\omega_{n'}
\langle n',\bm{k}'|
\tau,\bm{x}\rangle,
\quad
-i\partial_{i}
\langle n',\bm{k}'|
\tau,\bm{x}\rangle
=
-k^{'}_{i}
\langle n',\bm{k}'|
\tau,\bm{x}\rangle,
\end{align}
the Dirac operator acts as
\begin{align}
\mathcal{D}
\langle n',\bm{k}'|
\tau,\bm{x}\rangle
=
G^{-1}(i\omega_{n'},\bm{k}')
\langle n',\bm{k}'|
\tau,\bm{x}\rangle
\end{align}
with the inverse of the Green's function
$G^{-1}(i\omega_{n'},\bm{k}')
=
\gamma^0(i\omega_{n'}-\mu^*)
-\gamma^i k_i'
+m_{\mathrm D}$.
Thus, the eigenvalue of the Dirac operator
$\mathcal{D}$
is 
$G^{-1}(i\omega_{n'},\bm{k}')$.
Substituting this into the matrix element and using the orthonormality of the plane waves, we obtain
\begin{align}
\left\langle n,\bm{k}\left|
\mathcal{D}
\right|n',\bm{k}'\right\rangle
=
\delta_{n, n'}
\delta_{\bm{k}, \bm{k}'}
G^{-1}(i\omega_n,\bm{k}).
\end{align}
Consequently, the matrix elements of the logarithm becomes
\begin{align}
\left\langle n,\bm{k}\left|
\log\mathcal{D}
\right|n',\bm{k}'\right\rangle
=
\log
\left[
G^{-1}(i\omega_{n'},\bm{k}')
\right]
\delta_{n, n'}
\delta_{\bm{k}, \bm{k}'}.
\end{align}
Thus, 
\begin{align}
\mathrm{Tr}\log \mathcal{D}
=
\sum_{n,\bm{k}}
\mathrm{tr}_{\mathrm D}
\log
\left[
G^{-1}(i\omega_{n},\bm{k})
\right].
\end{align}
Finally, taking the thermodynamic limit,
\begin{align}
\sum_{\bm{k}}
\to
V\int\frac{d^3k}{(2\pi)^3},
\end{align}
we obtain the grand potential 
$\Omega$ 
as
\begin{align}
\Omega
=
-\frac{1}{\beta V}\log{Z}
=
\frac{1}{2}m^2_{\sigma}\sigma^2
-\frac{1}{2}m^2_{\omega}\omega^2
-\frac{1}{\beta}
\sum_{n}
\int \frac{d^3k}{(2\pi)^3}
\text{tr}
\log
\left[
G^{-1}(i\omega_{n}, \bm{k})
\right],
\end{align}
where we used
\begin{align}
\mathrm{Tr}\log \mathcal{D}
=
V
\sum_n
\int\frac{d^3k}{(2\pi)^3}
\mathrm{tr}_{\mathrm D}
\log
\left[
G^{-1}(i\omega_{n},\bm{k})
\right].
\end{align}
Up to an irrelevant constant independent of 
$T$
and
$\mu^*$, 
one finds~\cite{Cavagnoli2010}
\begin{align}
\mathrm{tr}_{D}
\left[
\log G^{-1}(i\omega_n,\bm{k})
\right]
=
2\log\left[
(i\omega_n-\mu^*)^2-\varepsilon_k^2
\right],
\end{align}
where
\[
\varepsilon_k=\sqrt{k^2+m_{\mathrm D}^2}.
\]
The fermionic contribution to the thermodynamic potential is therefore
\begin{align}
\Omega_{\text{F}}
=
-\frac{1}{\beta}
\sum_n
\int\frac{d^3k}{(2\pi)^3}
\mathrm{tr}_{\text{D}}
\log
\left[
G^{-1}(i\omega_n,\bm{k})
\right].
\end{align}
Performing the Matsubara sum gives
\begin{align}
\Omega_{\text{F}}
=
-2\int\frac{d^3k}{(2\pi)^3}\varepsilon_k
-\frac{2}{\beta}
\int\frac{d^3k}{(2\pi)^3}
\left[
\log\left(1+e^{-\beta(\varepsilon_k-\mu^*)}\right)
+
\log\left(1+e^{-\beta(\varepsilon_k+\mu^*)}\right)
\right].
\end{align}
The first term is the vacuum contribution. 
In the main text, we subtract this vacuum term and keep only the thermal and density-dependent part. 
Thus, the grand potential 
$\Omega$
becomes
\begin{align}
\Omega
=
\frac{1}{2}m_\sigma^2\sigma^2
-\frac{1}{2}m_\omega^2\omega^2
-\frac{2}{\beta}
\int\frac{d^3k}{(2\pi)^3}
\left[
\log\left(1+e^{-\beta(\varepsilon_k-\mu^*)}\right)
+
\log\left(1+e^{-\beta(\varepsilon_k+\mu^*)}\right)
\right].
\end{align}
Equivalently, introducing the Fermi-Dirac distribution function
\begin{align}
f_\mp(\varepsilon_k)
=
\frac{1}{e^{\beta(\varepsilon_k\mp\mu^*)}+1},
\end{align}
we finally obtain~\eqref{theremodynamicpotential}
\begin{align}
\Omega
=
\frac{1}{2}m_\sigma^2\sigma^2
-\frac{1}{2}m_\omega^2\omega^2
+
\frac{2}{\beta}
\int\frac{d^3k}{(2\pi)^3}
\left[
\log\left(1-f_-(\varepsilon_k)\right)
+
\log\left(1-f_+(\varepsilon_k)\right)
\right].
\nonumber
\end{align}

\end{appendix}

\bibliography{ref}

\end{document}